\documentclass[aps,prb,twocolumn]{revtex4-1}
\usepackage{graphicx}
\usepackage[bottom]{footmisc}
\usepackage{amsmath}
\usepackage{amssymb}
\usepackage{braket}
\usepackage{hyperref}
\usepackage{subfigure}
\usepackage[usenames,dvipsnames]{color}
\usepackage{graphicx}
\usepackage{subfigure}
\usepackage{cleveref}
\usepackage{amsmath,amssymb}

\newcommand{\ct}{\cite}

\newcommand{\bi}{\bibitem}
\newcommand{\be}{\begin{equation}}
\newcommand{\ee}{\end{equation}}
\newcommand{\ba}{\begin{eqnarray}}
\newcommand{\ea}{\end{eqnarray}}

\newcommand{\non}{\nonumber}

\begin{document}
\begin{@twocolumnfalse}

\title{Temporal variation in the winding number due to dynamical symmetry breaking and associated transport in a driven SSH chain}
\author{Souvik Bandyopadhyay, Utso Bhattacharya and Amit Dutta}
\affiliation{Department of Physics, Indian Institute of Technology, Kanpur, Kanpur 208016, India}

	\begin{abstract}
	Considering a BDI symmetric one-dimensional SSH model, we explore the fate of the bulk topological invariant, namely, the winding number under a generic  time dependent
	perturbation;  the effective Hamiltonian, that generates the temporal evolution of the initial (ground) state of the completely symmetric initial Hamiltonian, may have
	the same or different symmetries. To exemplify,
	we consider the following protocols, namely (i) a perfectly periodic protocol (ii) a periodic protocol with noisy perturbations and also (iii) sudden changes in the parameters of
	the initial Hamiltonian. We establish that the topological invariant may change in some cases when the effective Hamiltonian (or the Floquet Hamiltonian in the periodic situation
	when observed stroboscopically) does not respect all BDI symmetries; this is manifested in the associated particle  (polarisation) or heat current in the bulk.  Our results establish a strong connection  
	between the time evolution of the winding number (thus, the associated transport of currents)  and the symmetry of the Hamiltonian generating the time evolution which has been illustrated considering an exhaustive set
	of possibilities.

	\end{abstract}
	\maketitle
\end{@twocolumnfalse}
 
\section{Introduction}
\label{intro}
Recently, there has been an upsurge in the studies of topological condensed matter systems both from the theoretical\cite{kitaev01,kane05,bernevig06,fu08,zhang08,sato09,sau10a,sau10b,lutchyn10,oreg10} and experimental point of view \cite{mourik12,rokhinson12,deng12,das12,churchill13,finck13,alicea12,leijnse12,beenakker13,stanescu13} (For review see [\onlinecite{moore10,shen12,bernevig13}]). Topological phases of matter are characterised by gapped bulk states but with robust gapless excitations at the boundaries. Their novelty lies in the fact that they simply cannot be understood under the well established Landau-Ginzburg paradigm which classifies phases of matter in terms of spontaneous symmetry breaking. Two topological phases sharing the same symmetries may have very different global ordered ground states which cannot be adiabatically connected to each other without closing the gap in the bulk spectrum (i.e., without crossing a topological quantum critical point). \\

However, even in such systems symmetry considerations play a tremendous role. It has been possible to classify different topological phases of non-interacting systems \cite{altland97}  based on the constraints imposed on their Hamiltonians by the symmetries they obey; attempts are being made to achieve a similar classification for interacting systems \ct{kitaev11}. The lack of classification in terms of a local order parameter and the presence of a global (topological) order implies that the topological phases can be classified by determining certain bulk topological invariants such as the winding  or the Chern number \ct{shen12}, which depend on the global character of the eigenvectors representing the system.\\

Experimental realizations of topological systems especially in cold atomic setups\cite{bloch13,spielman15,jotzu14,bloch08,lewenstein12} have also opened up the possibility of subjecting them to time-dependent drives.
These experimental studies have initiated a plethora of theoretical works on quenched and periodically driven topological systems \cite{killi12,eckardt14,grushin16,kitagawa10,lindner11,bermudez09,patel13,thakurathi13,wang17,tarnowski19,liou18,jiang11,trif12,gomez12,dora12,cayssol13,liu13,tong13,katan13,kundu13,bastidas13,schmidt13,reynoso13,perez14}.  This enables us to probe the robustness of the topological features of the ground states of such systems against time-dependent perturbations. Moreover, time-dependent drives that are periodic in time also leads to the realisation of new topological phases of matter which have no equilibrium counterpart\cite{oka09,kitagawa11,rudner13,balseiro14,mitra15,gil16}. Interestingly, it has been shown that for two-dimensional (2D) Chern insulating systems without boundaries, it is not possible to reach a non-trivial topological state via unitary evolution from a trivial initial state as the bulk Chern number remains invariant\cite{rigol15}. However, for the same systems with boundaries the edge states of such 2D systems can exhibit non-trivial dynamics as the bulk-boundary correspondence for such systems in its usual form does not hold out of equilibrium\cite{cooper15,utso17}.\\

Recently, the fate of  topology in out of equilibrium one-dimensional (1D) situations  are being investigated along similar lines\cite{ginley18}. This brings us to the question that whether the out of equilibrium dynamics of 1D  topological systems also exhibit a behaviour similar to the 2D situation. To address this question,  we subject the paradigmatic topological 1D Su-Schrieffer-Heeger (SSH) model to time-dependent drives and  investigate the following questions: (a) Is it possible to change the winding invariants of such a system under the application of a time-periodic drive?  (b) Do the symmetry constraints of the time-dependent perturbations affect the topological properties of the post quench states?  
(c) What happens when the periodicity in time is not perfect but is affected by biased random noisy perturbations? Finally, (d) what happens to the energy transport dynamics of such systems under the application of non-equilibrium perturbations such as sudden quenches?\\

In this work, we initially focus on a  generically driven   SSH model  in which  we show that the bulk winding number charactarizing the 1D system do {\it vary} in time only when the applied time-dependent perturbations break certain symmetries of the un-driven Hamiltonian. This change of the winding number is also accompanied by the generation of an observable particle current; this  attains a steady value, asymptotically in time,  in the case of perfectly periodic driving.  However, we further show that the presence of a biased random noise in the periodic drive results in the generation of an infinite temperature state which is topologically featureless. In the noisy case, the accompanying particle current although settles to a pre-thermal region after exhibiting initial transient oscillations, but eventually  decays to zero asymptotically with time in accordance with the infinite temperature behaviour of the bulk invariant. Finally, we also focus on the possibility of local energy transport or heat current generation when the SSH system is subjected to sudden quenches. We show that the generation of heat currents in the system are related to different symmetry considerations of the applied time-dependent perturbations in comparison to   the production of particle currents in the same system.

The paper is organised in the following fashion: in Sec. \ref{sec_model}, we introduce the SSH model discussing the underlying topology and symmetry properties. The fate of the winding
number in a generic driven system is discussed in Sec. \ref{sec_winding_driven}. The special situation of the periodic driving is discussed in Sec. \ref{sec_periodic} where we show how
 the change in the winding number is manifested in the corresponding particle current generation in the bulk. Finally, in Sec. \ref{sec_heat_current}, we identify the symmetries of the effective Hamiltonian that result
in the heat current generation considering a sudden quenching protocol.

\section{The Su-Schrieffer-Heeger (SSH) Model}

\subsection{The topological transition}
\label{sec_model}
The SSH model\cite{shen12} which belongs to the BDI class of topological insulators is the simplest $1$D model exhibiting an underlying topological structure and end states [cite]. Physically, it describes a $1$D lattice with a two atom sublattice structure in which the intra-lattice hopping amplitude is in general different from the inter-lattice hopping amplitude. The Hamiltonian for the SSH model can be written in terms of the (spin polarised) fermion creation and annihilation operators as,
\begin{equation}\label{eq:H}
H=\sum_{n=1}^{N}(vc^{\dagger}_{n,1}c_{n,2}+wc^{\dagger}_{n,2}c_{n+1,1}+h.c),
\end{equation}
where {\it h.c}. denotes the hermitian conjugate and $v$ and $w$ are the intra-lattice and inter-lattice hopping amplitudes. The complex fermionic operator $c^{\dagger}_{n,i}$ ($c_{n,i}$) creates (destroys) a fermion in the sublattice position $i~(i=1,2)$ of the $n^{th}$ unit cell and satisfies the fermionic anti-commutation rules,
\begin{equation}\label{eq:anticom}
\{c^{\dagger}_p,c_q\}=\delta_{pq} \text{~and~} \{c_p,c_q\}=\{c^{\dagger}_p,c^{\dagger}_q\}=0.
\end{equation}
After performing a tight-binding analysis one can write the Hamiltonian in Eq.~\eqref{eq:H} as,
\begin{equation}\label{eq:Hk}
H(k)=\bigoplus_{k}\vec{h}(k).\vec{\sigma},
\end{equation} 
where, 
\begin{eqnarray}\label{eq:hk}
\nonumber h_{x}(k)&=& Re(v)+|w|\cos(k+arg(w))\\
\nonumber h_{y}(k)&=& -Im(v)+|w|\sin(k+arg(w))\\
\noindent h_{z}(k)&=& 0,
\end{eqnarray}
where the lattice parameter is set equal to identity.
This Hamiltonian has the following eigenvalue spectrum,
\begin{equation}\label{eq:Ek}
E(k)=\pm|\vec{h(k)}|,
\end{equation}
and the respective eigenvectors are,
\begin{equation}\label{eq:eg}
|\pm\rangle = \frac{1}{\sqrt{2}} \left(
\begin{array}{c}
\pm e^{-i \phi(k) } \\
1 \\
\end{array}
\right)
\end{equation}
where $\phi=\tan^{-1}\left(\frac{h_{y}}{h_{x}}\right)$.
It is clear from the Eq.~\eqref{eq:hk}, that $\vec{h}(k)$ is periodic in $k$ with a period of $2\pi$. Hence in the space of $h_{x}$ and $h_{y}$, $\vec{h}(k)$ traces out a closed curve as $k$ varies over the first Brillouin zone (in this case, a circle). Furthermore, the SSH model is classified through the following bulk  topological winding number $\nu$, which is given as, 
\begin{equation}\label{eq:winding}
\nu=\frac{i}{2\pi}\int^{\pi}_{-\pi}dk \frac{d}{dk}\ln(h_{x}+ih_{y})=\frac{i}{2\pi }\int_{BZ}\langle\psi_{0}^{k}|\partial_{k}|\psi_{0}^{k}\rangle dk,
\end{equation}
which is quantized and can only assume integral values and is proportional to the change in the argument of $\vec{h}(k)$ as $k$ varies over the first Brillouin zone. Hence, if the circle in the parameter space does not enclose the origin, $\nu$ is zero (i.e. if $v > w$). On the other hand,
$\nu$ is one if the circle encloses the origin (i.e. if $v < w$) and the chain hosts topologically protected robust end states. It is also evident from Eq.~\eqref{eq:Ek} that the energy gap between the two bands vanishes at $|v| = |w|$ and the winding number becomes undefined. Thus if the gap is not closed, $\nu$ is well defined and is robust to external changes in the Hamiltonian and hence is a topological invariant clearly demarcating the topologically trivial and non-trivial phases.\\

\subsection{Symmetries}~ The topological classification of non-interacting many-body quantum systems are performed by considering three different discrete symmetries viz., the time reversal symmetry ($\mathcal{T}$), the particle-hole symmetry ($\mathcal{P}$) and the sublattice (chiral) symmetry ($\mathcal{S}$). The constraints imposed upon the Hamiltonian of a system possessing the above symmetries in the quasi momentum basis are expressed as,
\begin{equation}\label{eq:symm}
\begin{split}
\mathcal{T}^{-1} H(k)\mathcal{T}=H(-k),\\
\mathcal{P}^{-1}H(k)\mathcal{P}=-H(-k),\\
\mathcal{S}^{-1}H(k)\mathcal{S}=-H(k),
\end{split}
\end{equation}
where $\mathcal{T}$ and $\mathcal{P}$ are anti-unitary operators such that $\mathcal{T}^2=\pm\mathbb{I}$ and $\mathcal{P}^2=\pm\mathbb{I}$ whereas $\mathcal{S}$ is an unitary operator satisfying where $\mathcal{S}^2=\mathbb{I}$ and $\mathbb{I}$ is the $2 \times 2$ identity operator. We also note that the sublattice symmetry is a combined effect of the time reversal symmetry and the particle hole symmetry as,
\begin{equation}\label{eq:sublattice}
\mathcal{S}=\mathcal{T}\mathcal{P}.
\end{equation}
It is now evident from the Hamiltonian of the SSH model in Eq.~\eqref{eq:Hk} and the symmetry transformations in Eq.~\eqref{eq:symm} that the SSH model is symmetric under the sublattice transformation $\mathcal{S}=\bigotimes_{k}\sigma_z$ which results in the vanishing of $h_z(k)$. Also, if the hopping coefficients $v$ and $w$ are real, the Hamiltonian possesses time reversal symmetry $\mathcal{T}=\bigotimes_k\mathcal{K}$, $\mathcal{K}$ being simply the complex conjugation operator. Hence, it is clear from Eq.~\eqref{eq:sublattice} that the system is also symmetric under the particle-hole/charge conjugation operation with $\mathcal{P}=\bigotimes_k\mathcal{K}\sigma_z$ Therefore, as such the SSH model belongs to the BDI class of  Hamiltonians within the topological classification scheme.\\

\section{The fate of winding number following a generic drive}
\label{sec_winding_driven}
We consider the SSH model and study the temporal evolution of the equilibrium topological invariant, i.e., the winding number under a generic unitary drive. We begin with an initial state $|\psi_{k}(0)\rangle$, the system is allowed to evolve under the driven Hamiltonian $H_k(t)$.  The state $|\psi_{k}(0)\rangle$ therefore evolves with time as
\begin{equation}\label{eq:evolve}
\begin{split}
|\psi_{k}(t)\rangle=\mathbb{T} e^{-i\int_{0}^{t}H_{k} (t^{'})dt^{'}}|\psi_{k}(0)\rangle\\ 
\equiv e^{-iH_{k}^{\rm eff}(t) t}|\psi_{k}(0)\rangle\\
=U_{k}(t)|\psi_{k}(0)\rangle
\end{split}
\end{equation}
where $H^{\rm eff}_k(t)$ is the time-dependent effective Hamiltonian acting as a generator of the unitary evolution acting on  the driven system and $\mathbb{T}$ denotes the time ordering operator. We now investigate the fate of the winding number under such a time-dependent dynamics. To analyse this,  let us recall the time-dependent or dynamical Berry connection as
\begin{equation}\label{eq:db}
A_{k}(t)\equiv\left[\langle\psi_{k}(0)|U_k^{\dagger}\right]\partial_{k}\left[ U_{k}|\psi_{k}(0)\rangle \right],
\end{equation}
which evolves in time as,
\begin{eqnarray}\label{eq:conn}
A_{k}(t)&=&\langle\psi_{k}(0)|\partial_{k}|\psi_{k}(0)\rangle + \langle\psi_{k}(0)|U_k^{\dagger}(\partial_{k}U_{k})|\psi_{k}(0)\rangle \non \\
&=& A_{k}(0)+ \langle\psi_{k}(0)|U_k^{\dagger}(\partial_{k}U_{k})|\psi_{k}(0)\rangle.
\end{eqnarray}
Hence,  the change in the Berry connection at a later time is given by,
\begin{equation}
\Delta A_{k}=A_{k}(t)-A_{k}(0)=\langle\psi_{k}(0)|U_k^{\dagger}(\partial_{k}U_{k})|\psi_{k}(0)\rangle
\label{eq_connection}
\end{equation}
Recasting the effective Hamiltonian to the following form, $H_{k}^{\rm eff}(t)=|m(k,t)|\left(\hat{m}(k,t).\vec{\sigma}\right)$ and
also denoting $|m(k,t)|$ simply as $m$ we obtain,
\begin{eqnarray}\label{eqn_udu}
 U_k^{\dagger}(\partial_{k}&U_{k}&)=\partial_{k}m\{-it\sin^2{mt}(\hat{m}\cdot\sigma) \nonumber\\
 &+&i\sin^2{mt}(\hat{m}\times\partial_{k}\hat{m})\cdot\sigma\}-i(\sin{mt}\cos{mt})\partial_{k}\hat{m}\cdot\sigma \nonumber\\
\end{eqnarray}

The initial state $|\psi_{k}(0)\rangle$  that we consider happens to be the ground state of the SSH Hamiltonian (belonging BDI class) which  can be chosen to be of the form of Eq.~\eqref{eq:eg} where $\phi(k)$ is an odd function of $k$. Interestingly, the terms on the right hand side of the  Eq.~\eqref{eqn_udu} can be shown to vanish individually when integrated over the entire Brillouin zone, pertaining to certain conditions imposed upon the effective Hamiltonian $H^{\rm eff}_{k}$ as discussed below.\\

Let us now analyse the implications of Eqs.~ \eqref{eq_connection} and \eqref{eqn_udu}. Taking the expectation value of the first term of the above equation with respect to the state $|\psi_{k}(0)\rangle$, one observes 
that the integral of this quantity over the full Brillouin zone vanishes identically if $m_{x}(k)$ is an even function of $k$ and $m_{y}(k)$ is an odd function of $k$. Similarly, analysing the integral of the next two terms over the full Brillouin zone, we see that both of them vanishes identically if $m_{z}(k)$ is an odd function of $k$ or zero in addition to the above constraints imposed on $m_{x}(k)$ and $m_{y}(k)$. If the above conditions are satisfied by the effective Hamiltonian then the winding number must remain invariant in time.\\

It is evident from Eq.~\eqref{eq:symm} that the above constraints on the single particle Hamiltonian in $k$ space, demand the presence of certain symmetries of the effective Hamiltonian.
Namely, one concludes that the equilibrium winding number remains invariant under temporal evolution if the effective dynamical Hamiltonian $(H^{\rm eff}_{k})$ respects either of the symmetry combinations, $\mathcal{T}$ and $\mathcal{P}$ ($m_x(k)\rightarrow$ even, $m_y(k)\rightarrow$ odd, $m_z(k)\rightarrow 0$) or just $\mathcal{P}$ ($m_x(k)\rightarrow$ even, $m_y(k)\rightarrow$ odd, $m_z(k)\rightarrow$ odd).\\


\section{Periodic driving and non-equilibrium current generation}

\label{sec_periodic}

\subsection{Generic periodic driving}

\label{sec_periodic_generic}

\begin{figure*}
	\begin{center}
		\subfigure[]{\label{1a}}{\centering\includegraphics[width=0.47\textwidth,height=0.65\columnwidth]{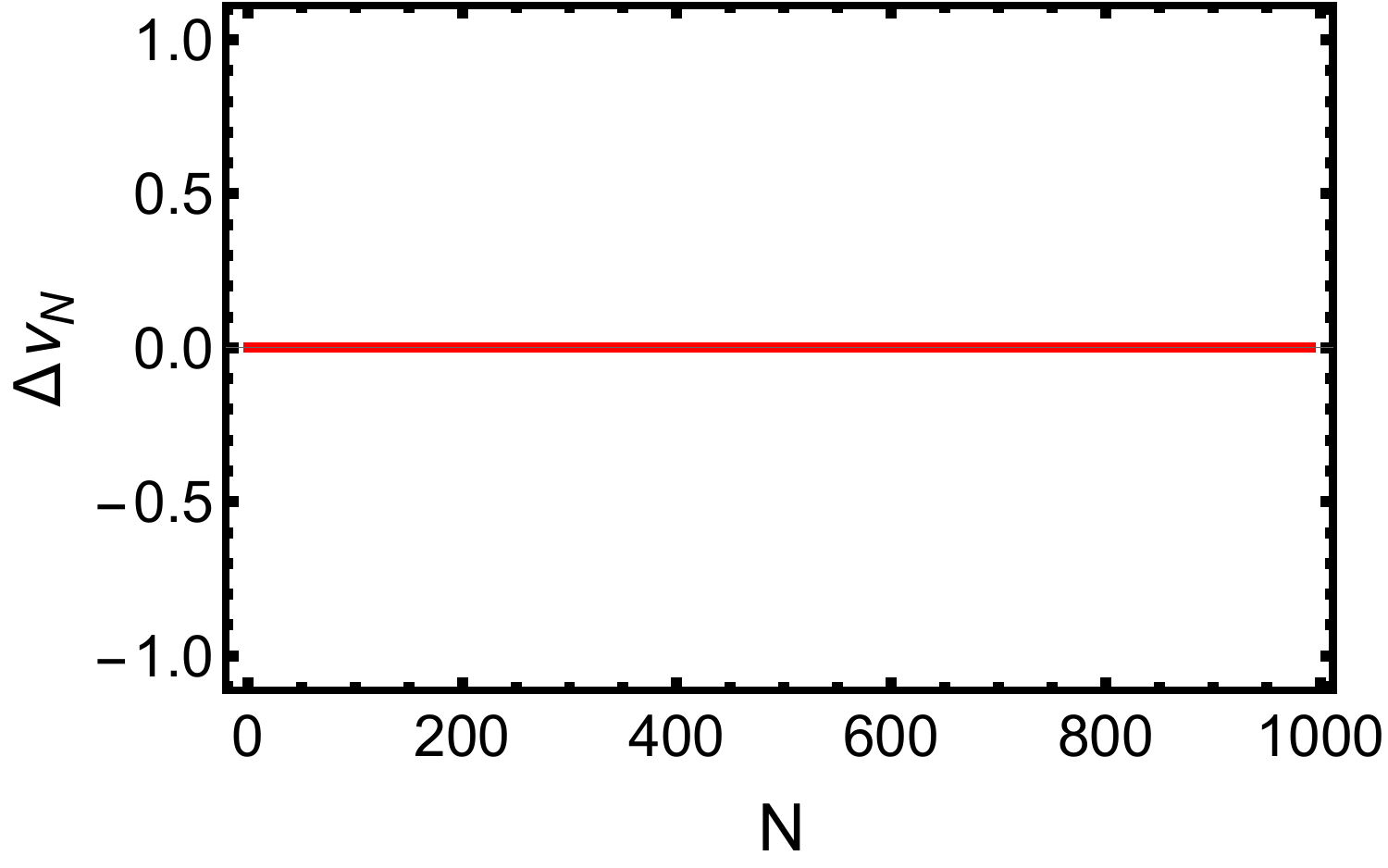}}
		\hspace{0.7cm}\subfigure[]{\label{1b}}\quad{\includegraphics[width=0.46\textwidth,height=0.65\columnwidth]{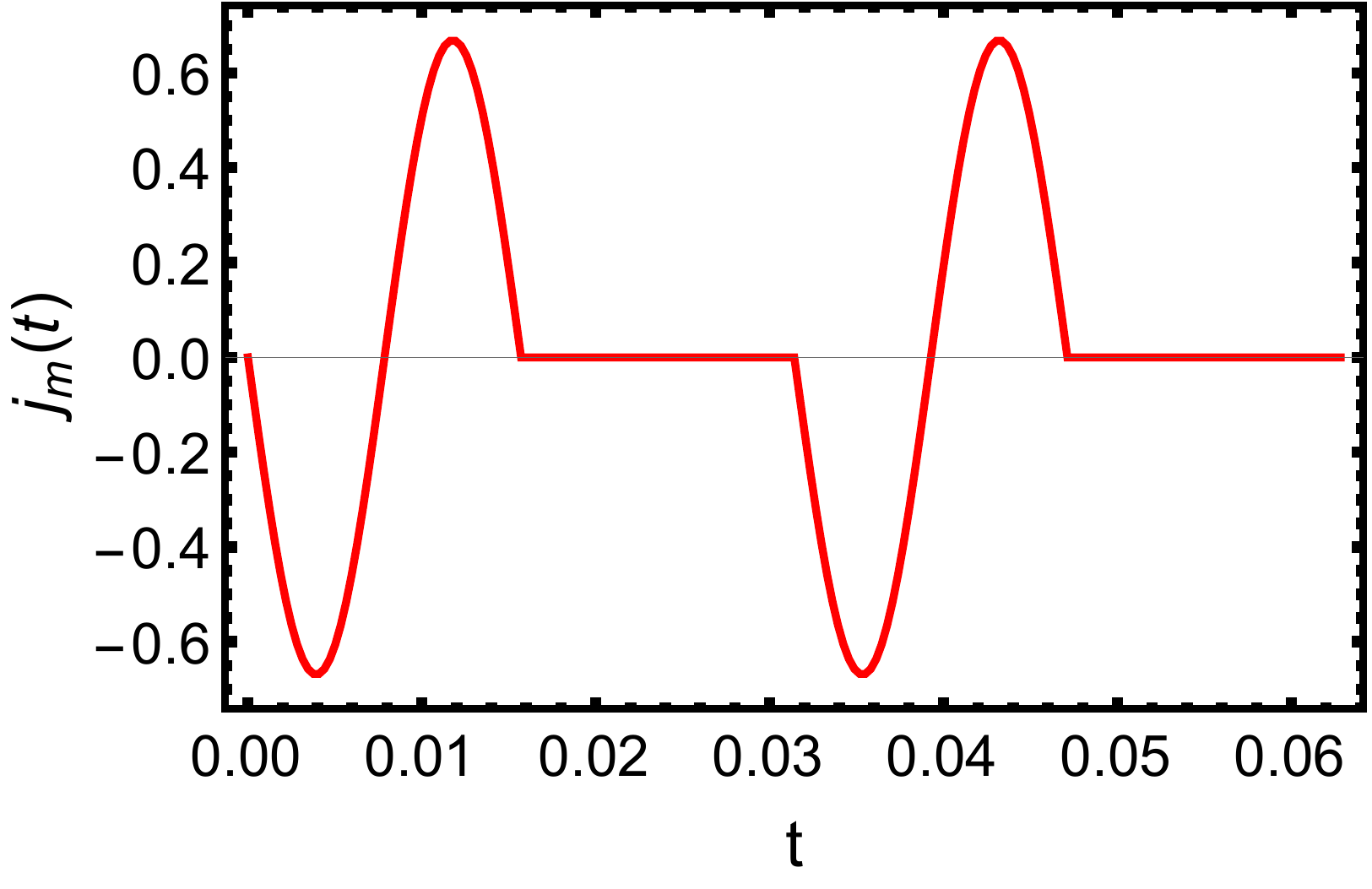}}
		
		\caption{(a) Stroboscopic change in winding number under a periodic drive that breaks only the  $\mathcal{P}$ symmetry in the time varying Hamiltonian $H_k(t)$ and the effective Hamiltonian $H^{\rm eff}_{k}$ while preserving all symmetries in the Floquet Hamiltonian, which is a BDI SSH chain with the hopping parameters $v=0.6$ and $w=0.8$ for a system size of $L=1000$.
			(b)  Particle current generation in micromotion within a time interval $\left(0,T\right)$ under a periodic drive breaking $\mathcal{P}$ symmetry in the time varying Hamiltonian $H_k(t)$ and the effective Hamiltonian $H^{\rm eff}_{k}$ while preserving all symmetries in the Floquet Hamiltonian, which is gain a BDI SSH chain with the hopping parameters $v=0.6$ and $w=0.8$ for a system size of $L=1000$. These cases are discussed in Sec.~ \ref{sec_periodic_generic}.  }
	\end{center}
\end{figure*}

In this section, our focus is to look at an observable which is the bulk polarization current density for various periodic driving protocols. It is straightforward to show \ct{ginley18} that in an arbitrary time-dependent situation the bulk polarization current density $j(t)$ of the SSH chain is directly proportional to the rate of change of the topological winding number $(\nu)$:
\begin{equation}\label{w_j}
j(t)=\frac{1}{2\pi}\int_{BZ}\bra{\psi_k(t)}\partial_{k}H_k(t)\ket{\psi_k(t)}=\frac{d\nu}{dt}
\end{equation}
where $\ket{\psi_k(t)}$ is the time evolved state for each quasi-momenta mode $k$ and $H_k(t)$ is the instantaneous time-dependent Hamiltonian.
However, in the case of a time periodic drive with a period $T$, the stroboscopic (measured after a complete period), the variation of the winding number denoted as $\Delta\nu_{m}$ for the $m-$th stroboscopic interval, is related to the average change in the bulk polarization density of the chain within the $(m-1)-$th and the $m-$th period of evolution i.e.,
\begin{equation}\label{avgJ}
\Delta\nu_{m}=\frac{\nu(mT)-\nu((m-1)T)}{T}=\frac{1}{T}\int_{(m-1)T}^{mT}j(t)dt
\end{equation}
The average polarization over one time period if expanded shows a dependency on the symmetries of both the effective Hamiltonian $(H^{\rm eff}_{k}(t))$ and the instantaneous time dependent Hamiltonian $(H_k(t))$,
\begin{equation}\label{avgJ1}
\begin{split}
\Delta\nu_m=\int_{(m-1)T}^{mT}\int_{BZ}dtdk\bra{\psi_k(0)}e^{iH^{\rm eff}_k(t)t}\\
\partial_{k}H_k(t)e^{-iH^{\rm eff}_{k}(t)t}\ket{\psi_k(0)}
\end{split}
\end{equation}
Thus, if both $H^{\rm eff}_{k}(t)$ defined at every instant $t$ but lying within the stroboscopic interval and the instantaneous $H_k(t)$ preserve the above symmetries enlisted in Sec. \ref{sec_winding_driven}, $\Delta\nu_m$ is zero at every instant of time $t$. Consequently the stroboscopic winding number $\nu(mT)$ remains trivially invariant under the dynamics. On the contrary, if the Floquet Hamiltonian defined 
over a complete period as
$$ \exp (-i H_F (k) T) = \mathbb{T} e^{-i\int_{0}^{T}H_{k} (t^{'})dt^{'}}, $$
 preserves the identified symmetries while the effective Hamiltonian and the time dependent Hamiltonian breaks the symmetries explicitly within the time interval, the winding number is seen to remain invariant stroboscopically but not within the time period of the drive.  Therefore, although an instantaneous polarization current gets generated but when averaged over a complete period, it is seen to vanish despite being finite within a period of the cycle (Fig.~\ref{1a} and Fig.~\ref{1b}). This becomes evident when starting from the eigenstate of a completely $\mathcal{T}$,   $\mathcal{P}$ and $\mathcal{S}$ symmetric SSH model, using Eq.~\eqref{eq_connection}, the stroboscopic change in the winding number is observed to depend only on the symmetries of the Floquet Hamiltonian, i.e.,
\begin{equation}\label{delta nu}
\nu(mT)=\nu(0)+\frac{i}{2\pi}\oint\bra{\psi_k(mT)}(\partial_{k}e^{-iH_F(k)mT})\ket{\psi_k(0)}dk
\end{equation}
Thus, if $H_F(k)$ respects either $\mathcal{P}$ or $\mathcal{P}$ and $\mathcal{T}$ symmetries, the stroboscopic winding number remains dynamically invariant. We therefore conclude
that it is the symmetry of $H_F$ that determines the stroboscopic variation of the winding number.\\

\begin{figure*}
	\begin{center}
		\subfigure[]{\label{2a}}{\includegraphics[width=0.46\textwidth,height=0.63\columnwidth]{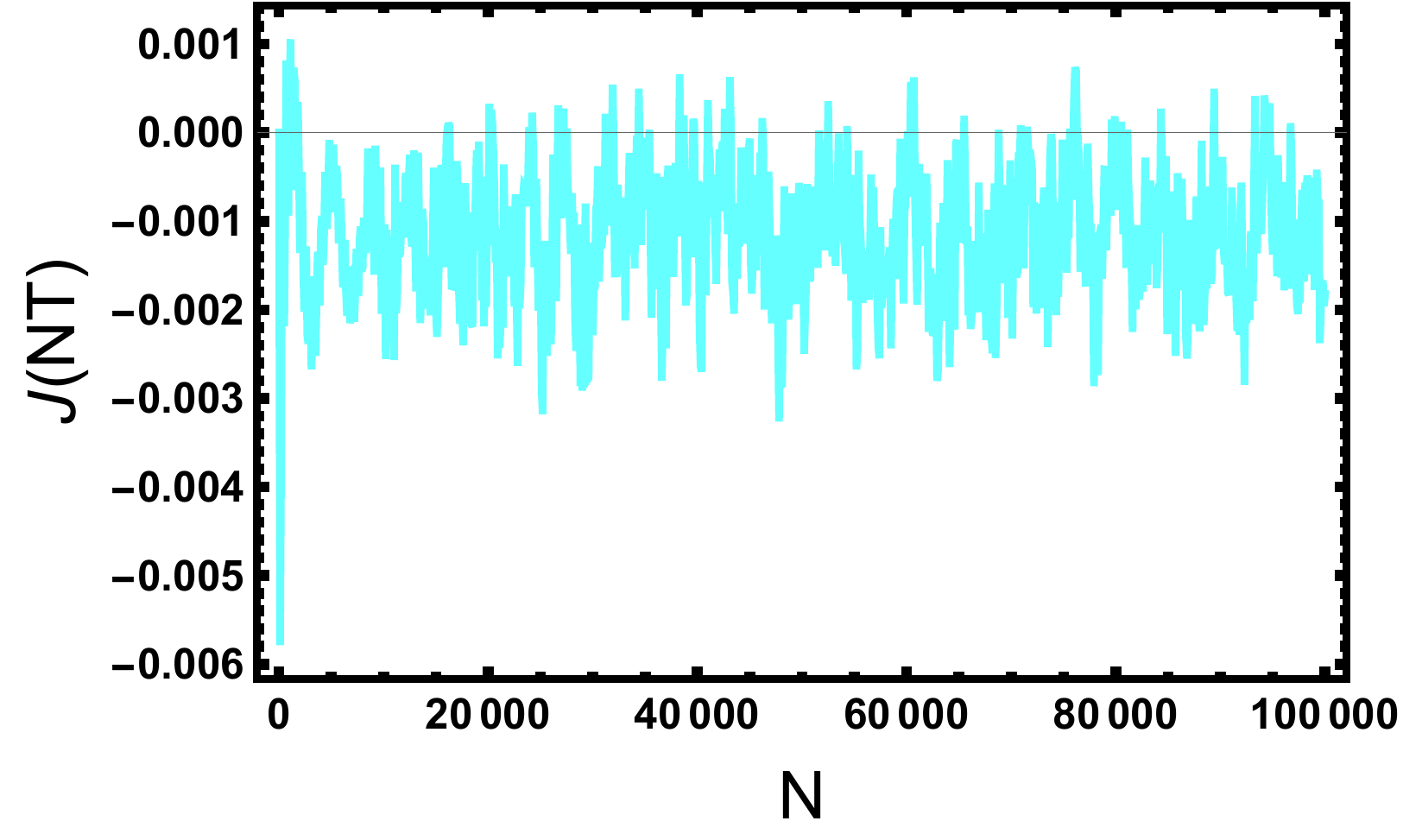}}
		\hspace{0.5cm}\subfigure[]{\label{2b}}\quad{\includegraphics[width=0.46\textwidth,height=0.63\columnwidth]{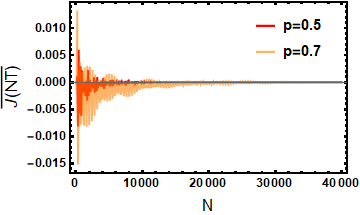}}
		\hspace{0.1cm}\subfigure[]{\label{2c}}{\includegraphics[width=0.46\textwidth,height=0.65\columnwidth]{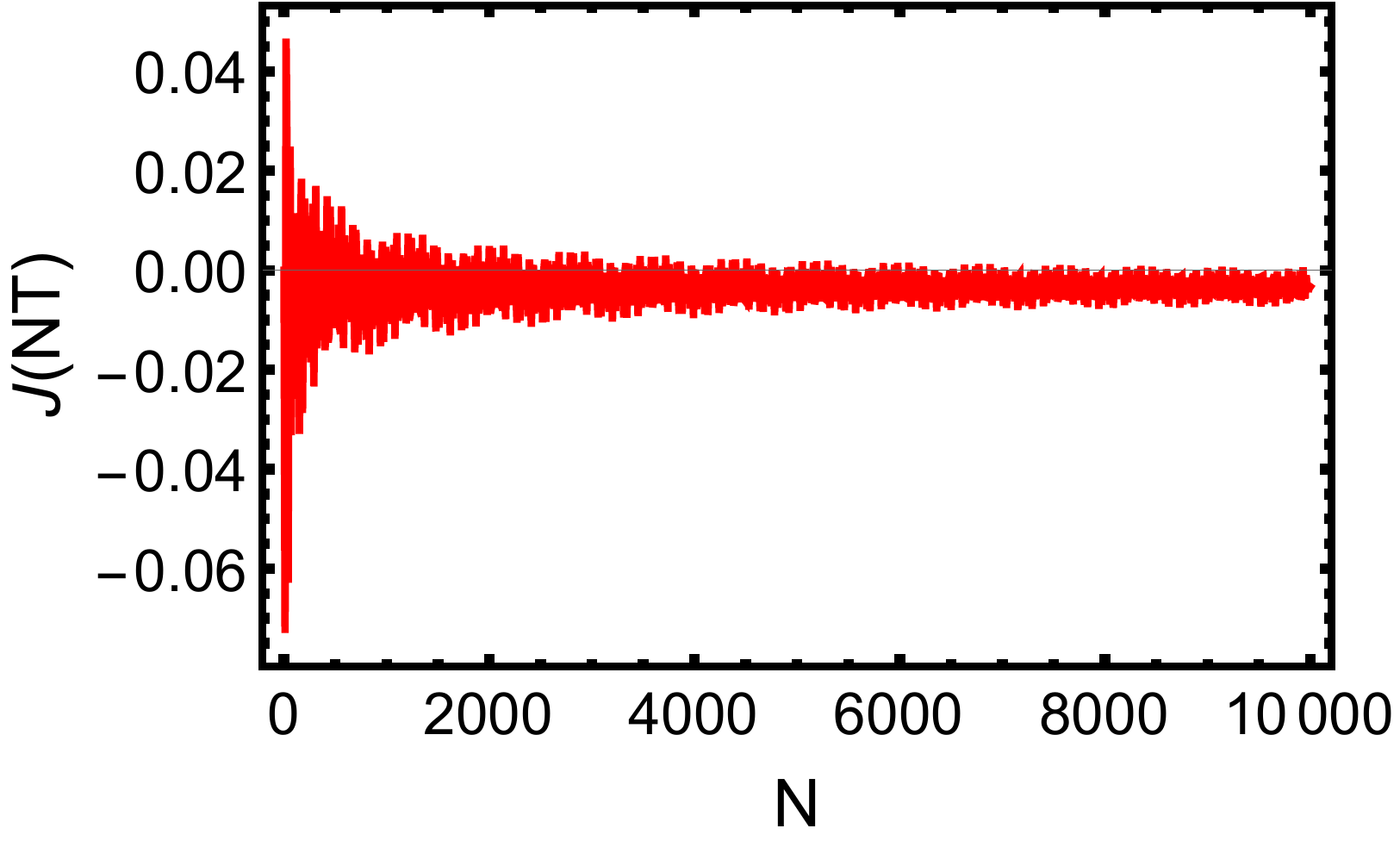}}
		\hspace{0.9cm}\subfigure[]{\label{2d}}{\includegraphics[width=0.46\textwidth,height=0.63\columnwidth]{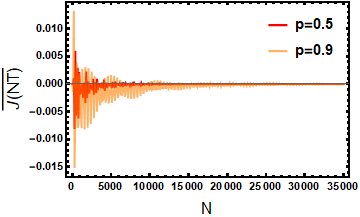}}
		\caption{ (a) Stroboscopic particle current in a periodically driven SSH chain, breaking $\mathcal{P}$ symmetry  by introducing a real NNN hopping in $H_1(k)$  (as discussed in \ref{sec_breaking_P_symmetry})  while preserving $\mathcal{T}$ in the Floquet Hamiltonian. The initial and final hopping strengths are chosen to be $v=2.5$, $w=1.5$ and a NNN hopping strength of $f=1.0$ for a system size $L=500$; the frequency of the periodic drive being $\omega=100$.
			(b)   Disordered averaged stroboscopic particle current in the corresponding aperiodic situation (as discussed in Sec.~\ref{sec_aperiodic}).
						(c) Stroboscopic particle current in a periodically driven SSH chain, breaking $\mathcal{P}$  by introducing a staggered mass in $H_1(k)$ (as discussed in \ref{sec_breaking_P_symmetry}) while preserving $\mathcal{T}$ in the Floquet Hamiltonian. The initial and final hopping strengths are chosen to be $v=0.2$, $w=1.5$ and a stagerred mass of $M=1.0$ for a system size $L=1000$,  and $\omega=100$.
			(d) The disordered averaged stroboscopic particle current in the corresponding aperiodic situation.
			 		}
	\end{center}
\end{figure*}

\begin{figure*}
	\begin{center}
		\subfigure[]{\label{3a}}\quad{\includegraphics[width=0.49\textwidth,height=0.65\columnwidth]{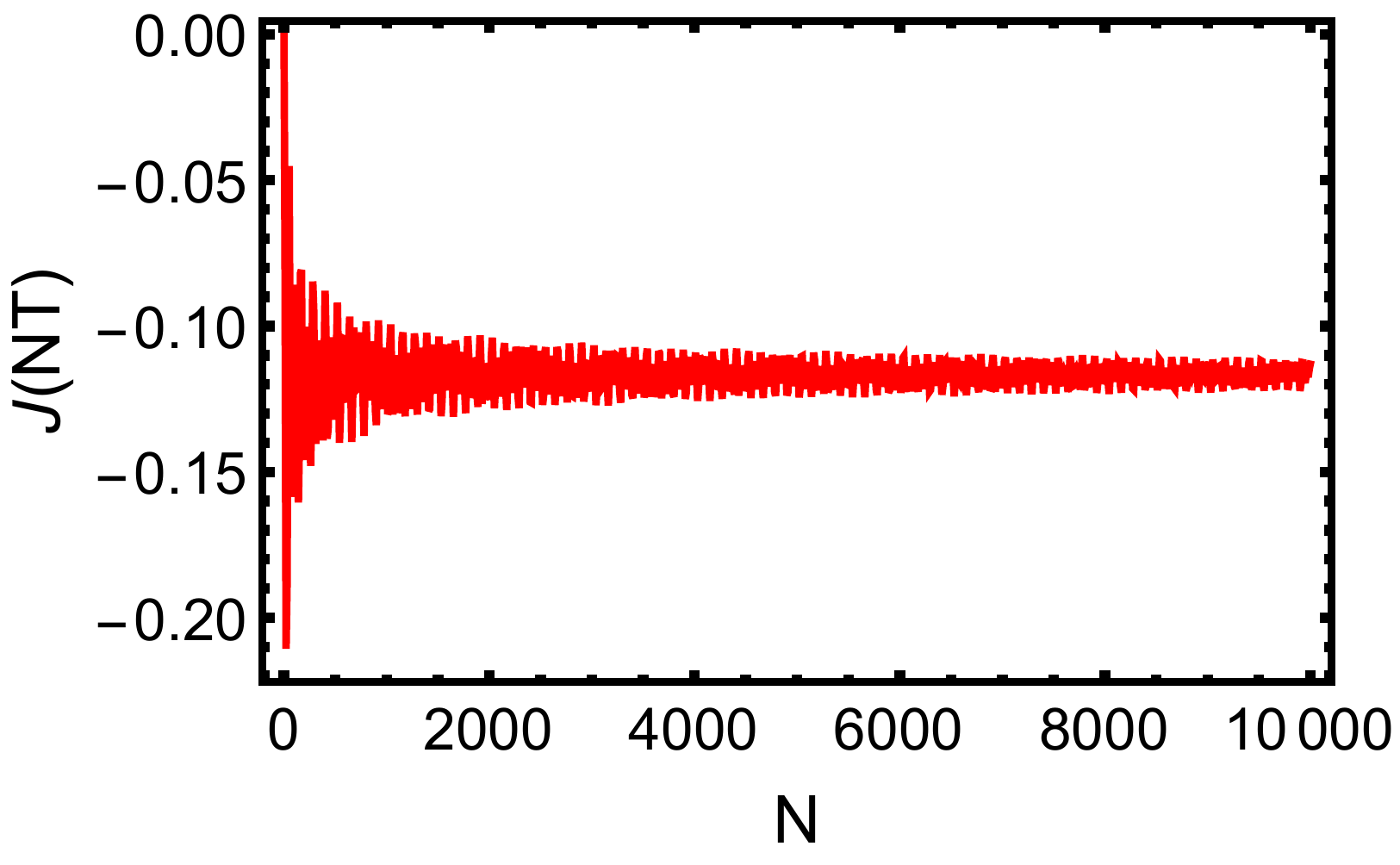}}
		\hspace{0.1cm}\subfigure[]{\label{3b}}{\includegraphics[width=0.48\textwidth,height=0.63\columnwidth]{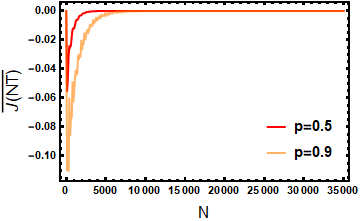}}
		\subfigure[]{\label{3c}}{\includegraphics[width=0.50\textwidth,height=0.65\columnwidth]{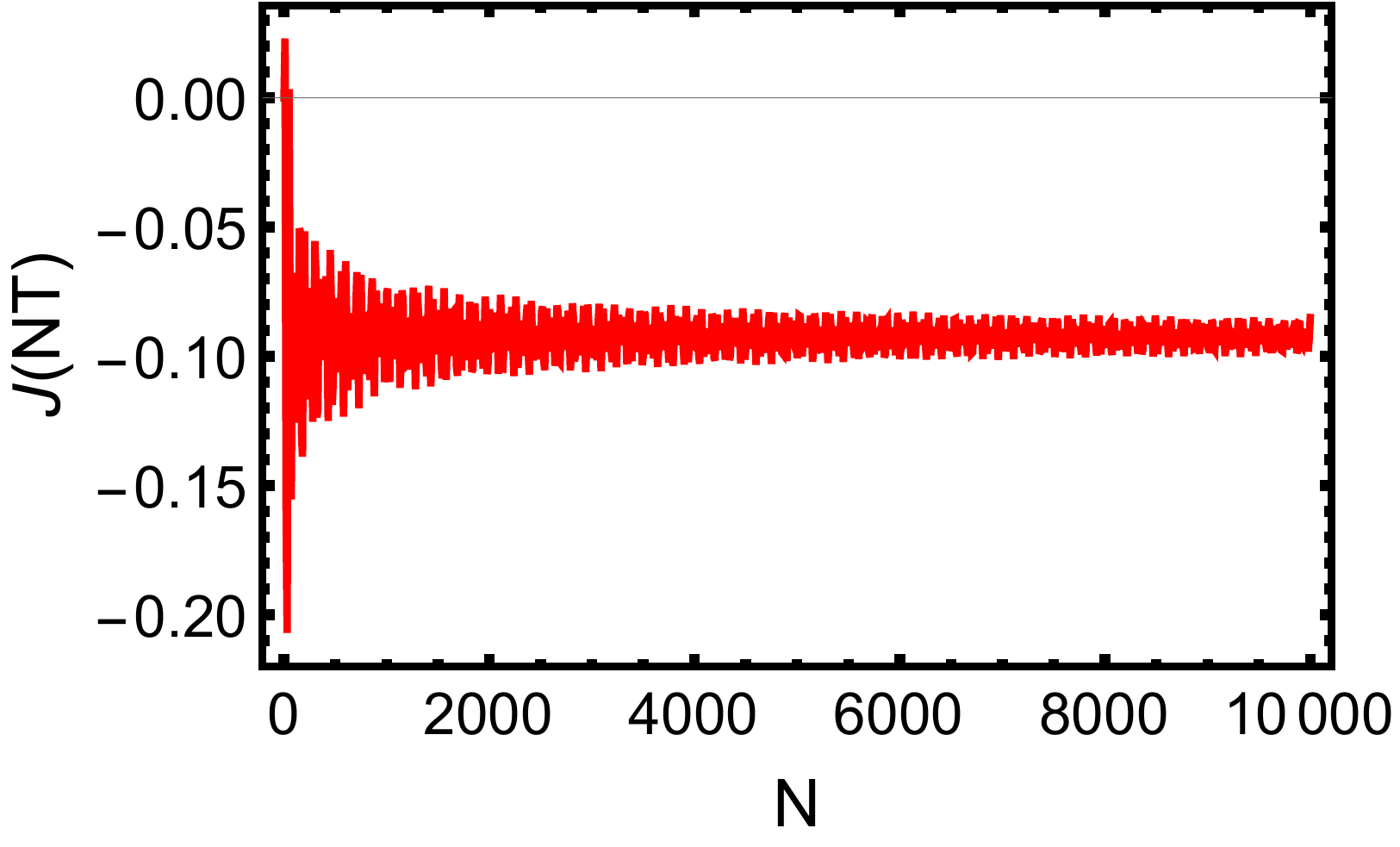}}
		\hspace{0.1cm}\subfigure[]{\label{3d}}{\centering\includegraphics[width=0.48\textwidth,height=0.65\columnwidth]{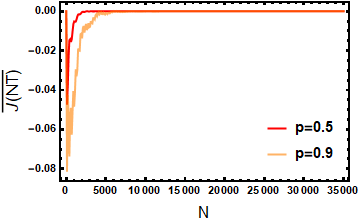}}

		\caption{(a) Stroboscopic particle current in a periodically driven SSH chain, breaking $\mathcal{P}$ and $\mathcal{T}$ by introducing an imaginary inter-cell hopping in $H_1(k)$ but preserving $\mathcal{S}$ in the Floquet Hamiltonian (as discussed in Sec.~\ref{sec_breaking_PandT}). The initial and final hopping strengths are chosen to be $v=0.2$, $w=1.5i$ for a system size $L=1000$,  with $\omega=100$.
			(b) Disordered averaged stroboscopic particle current in the corresponding aperiodic situation (as discussed in Sec.~\ref{sec_aperiodic}).
			(c) Stroboscopic particle current in a periodically driven SSH chain, breaking $\mathcal{P}$, $\mathcal{T}$  and $\mathcal{S}$ by introducing a staggered mass and an imaginary inter-cell hopping in $H_1(k)$  (as discussed in Sec.~\ref{sec_breaking_PTS}). The initial and final hopping strengths are chosen to be $v=0.2$, $w=1.5$ and a staggered mass of $M=1.0$ for a system size $L=1000$, $\omega=100$.
			(d) Disordered averaged stroboscopic particle current in the corresponding aperiodic situation.	
			}
	\end{center}
\end{figure*}

\begin{figure*}
	\begin{center}
		\hspace{0.1cm}\subfigure[]{\label{4a}}{\includegraphics[width=0.50\textwidth,height=0.65\columnwidth]{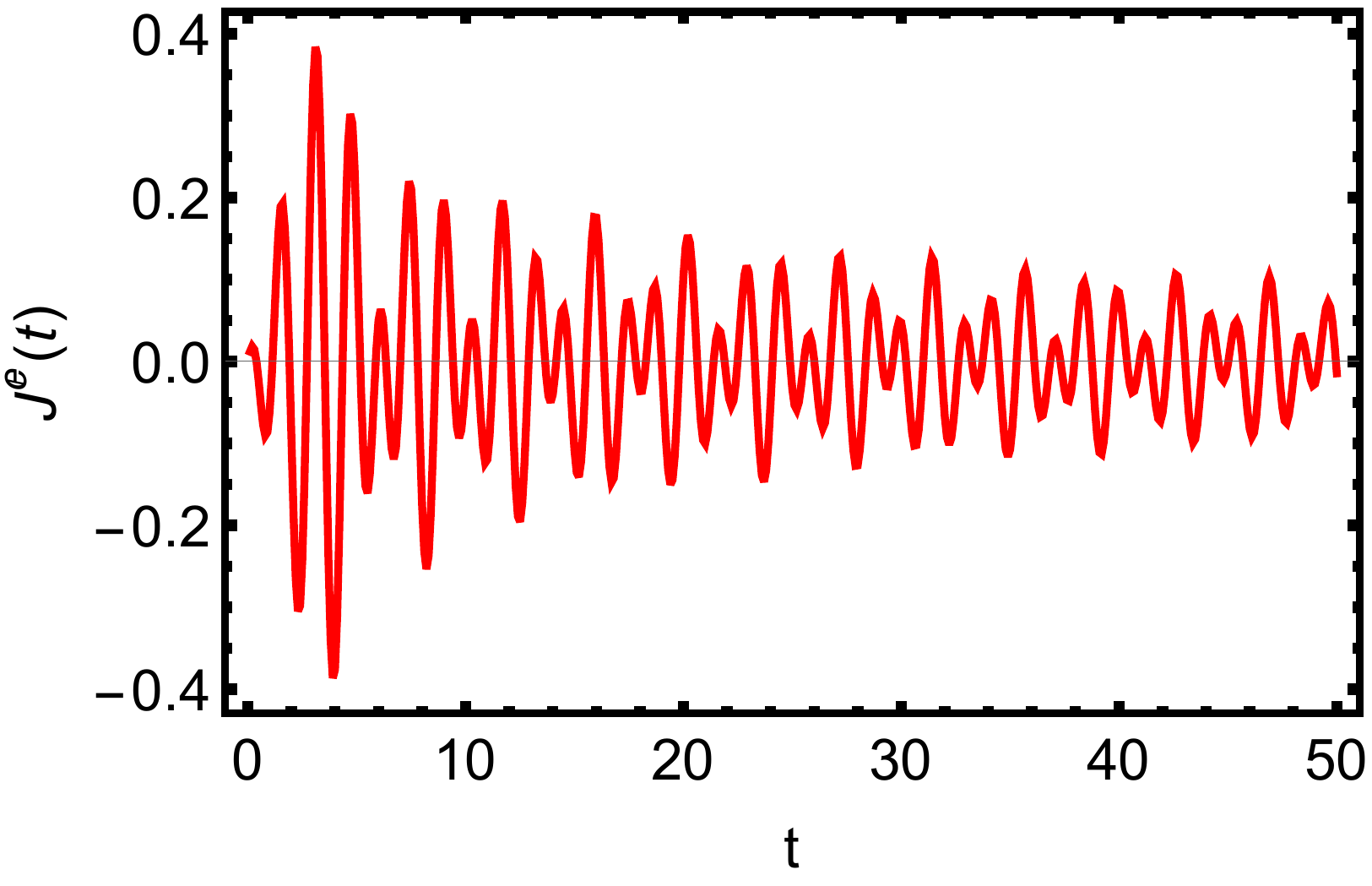}}
		\subfigure[]{\label{4b}}{\centering\includegraphics[width=0.48\textwidth,height=0.65\columnwidth]{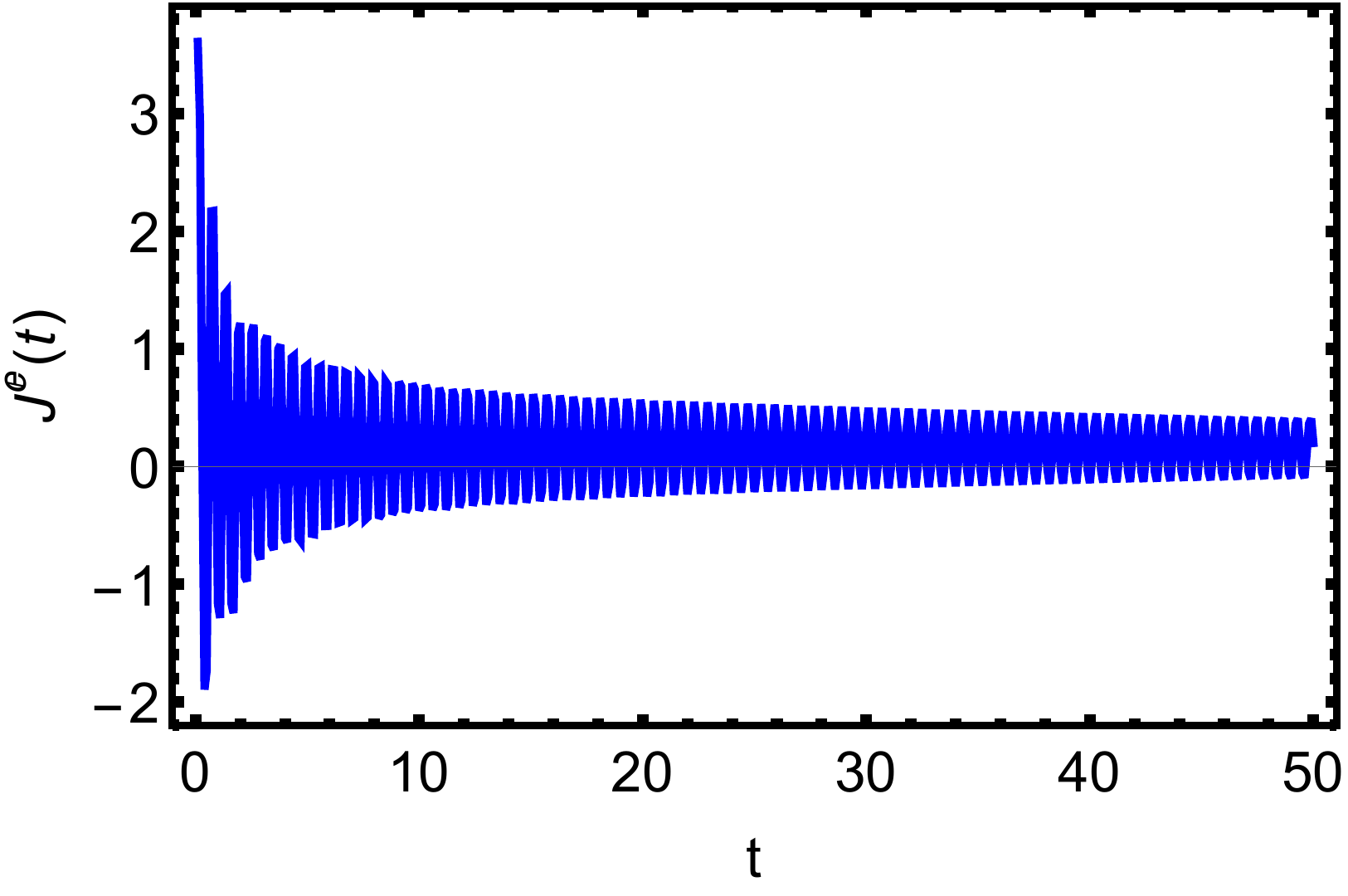}}

		\caption{(a) Heat current generation in the bulk in a quenched SSH chain by breaking $\mathcal{P}$ through the  sudden introduction of a staggered mass  which  preserves $\mathcal{T}$ in the  final quenched Hamiltonian  ${\cal H}(k)$ (as discussed in Sec.~\ref{sec_heat_breaking_P}). The initial and final hopping strengths are chosen to be $v=0.5$, $w=1.5$ and a staggered mass in ${\cal H}(k)$ of $M=1.0$ with $L=500$.
			(b) Heat current generation in the bulk in a quenched SSH chain by breaking $\mathcal{T}$ through the sudden introduction of an imaginary staggered NNN hopping in ${\cal H}(k)$ while preserving $\mathcal{P}$ in the final quenched Hamiltonian (as discussed in Sec.~\ref{sec_heat_breaking_T}). The initial and final hopping strengths are chosen to be $v=0.5$, $w=1.5$ and a NNN hopping strength of $f=5.0i$ in ${\cal H}(k)$ and  $L=500$.}
	\end{center}
\end{figure*}

\subsection{Explicit symmetry breaking in a multi-step periodic drive}

In this section, we will be considering different kinds of  periodic drives on a zero current carrying initial state of the SSH model, to probe whether after an asymptotically long time, the bulk polarization current density generated  due to the change in the winding invariant, attains a steady value when observed stroboscopically.

We consider a periodic drive  with a two step driving protocol applied within one stroboscopic time interval $(0,T)$;  on explicitly breaking certain symmetries in the Floquet Hamiltonian, the steady state current attains a constant value starting from a zero current initial state. Hence, a steady current can be {\it generated} in 1D topological systems by periodically driving provided that one dynamically breaks certain symmetries to be discussed below. 

To achieve the symmetry breaking, we employ the two-step periodic drive which involves the evolution of the initial state of the system under two piece-wise continuous time-independent Hamiltonians {\it viz.} $H_0(k)$ ($\in BDI$) and $H_1(k)$ in alternate time steps of width $T/2$. 
The effective propagator after the two time steps or after one complete period of driving therefore assumes the following form:
\begin{equation}\label{eq: periodic1}
U_k(T)=e^{-iH_1(k)\frac{T}{2}}e^{-iH_0(k)\frac{T}{2}}
\end{equation}

\subsection{Breaking $\mathcal{P}$ but preserving $\mathcal{T}$ and $\mathcal{S}$}

\label{sec_breaking_P_symmetry}
The breaking of the $\mathcal{P}$ symmetry in Eq.~\eqref{eq: periodic1} can be achieved in a variety of ways which we illustrate below: 
 (i)  by introducing  a real staggered next nearest neighbour hopping ($B_1$) and  (ii) by adding a staggered on-site potential ($B_2$). Referring to \eqref{eq:hk},  the Hamiltonian $H_1(k)$ occuring in Eq.~\eqref{eq: periodic1} for these situations are as follows,\\\\
{\it $(B_1)$} Staggered NNN hopping:\\
\begin{equation}\label{eq: B1}
H_1(k)=(v+w\cos{k})\sigma_x+w\sin{k}\sigma_y+f\cos{k}\sigma_z
\end{equation}
{\it $(B_2)$ Staggered on-site potential:}\\
\begin{equation}\label{eq: B2}
H_1(k)=(v+w\cos{k})\sigma_x+w\sin{k}\sigma_y+M\sigma_z
\end{equation}

The bulk polarization current defined in Eq.~\eqref{w_j} in the situations $B_1$ and $B_2$ reaches a steady value starting from zero; this is shown Fig.~\ref{2a} and Fig.~\ref{2c}, respectively. In both the cases however, the time reversal symmetry remains preserved.\\ 

 \subsection{Breaking $\mathcal{P}$ and $\mathcal{T}$ But Preserving $\mathcal{S}$}
 \label{sec_breaking_PandT}
 
 We now break the $\mathcal{T}$ and $\mathcal{P}$ symmetry together in $H_F$,  by making the inter-cell hopping amplitude $w$ to be completely imaginary in $H_1(k)$ in Eq.~\eqref{eq: periodic1}. This however preserves the chiral symmetry $(\mathcal{S})$ in Eq.~\eqref{eq: periodic1} resulting in
 \begin{equation}\label{eq: allB}
 H_1(k)=(v - w\sin{k})\sigma_x + w\cos{k}\sigma_y
 \end{equation}
 The bulk polarization current defined in Eq.~\eqref{w_j} in this scenario also reaches a steady value starting from zero as can be observed from Fig.~\ref{3a}.  
 \subsection{Breaking $\mathcal{P}$, $\mathcal{T}$ and $\mathcal{S}$}
 \label{sec_breaking_PTS}
 
 Furthermore, the breaking of all the three symmetries $\mathcal{P}$, $\mathcal{T}$ and $\mathcal{S}$ can simply be achieved by selecting,
 \begin{equation}\label{eq: B2}
 H_1(k)=(v-w\sin{k})\sigma_x+w\cos{k}\sigma_y+M\sigma_z
 \end{equation}
 where $M$ is a constant quasi-momenta independent mass term. The bulk polarization current once again attains a steady non-zero value as can be seen in Fig.~\ref{3c}.

\section{Bulk topological invariant in an aperiodically driven system}

\label{sec_aperiodic}

We consider two 1D SSH Hamiltonians, one in the BDI class i.e., respecting both time reversal and particle hole symmetries whereas the other breaks either or both of the symmetries $\mathcal{P}$ and $\mathcal{T}$ along with  $\mathcal{P}$. Thus, under the first Hamiltonian $H_0(k)$ the winding number remains invariant in time, whereas under the second Hamiltonian $H_1(k)$ the winding number does not. The system is then subjected to an imperfect drive with an inherent probalistic evolution where at each time step of width $T/2$ the the dynamical Hamiltonian is chosen randomly between $H_0(k)$ and $H_1(k)$. The choice between $H_0(k)$ and $H_1(k)$ depends on the value of a random variable following a binomial distribution with a bias $p$. In the context of thermalization, such a random driving protocol have been studied in Refs. [\onlinecite{nandy17,utso18}]. \\
Introducing aperiodicity in the driving protocol, renders the system dynamically non-integrable. Under an aperiodic drive, observing the dynamics stroboscopically at intervals of $T$ modifies the relation established in Eq.~\eqref{avgJ} connecting the non-equilibrium stroboscopic current density and the temporal evolution of the bulk topological index. The current density when averaged over all disorder configurations and over a complete period $T$ results in,

\begin{equation}\label{eq:apcurrent}
\begin{split}
\frac{1}{T}\int_{(m-1)T}^{mT}dt~\overline{ j(t)}=\frac{1}{T}\left[\overline{\nu(mT)}-\overline{\nu((m-1)T)}\right],
\end{split}
\end{equation}
where $|\psi_k(mT)\rangle=\prod_{n=1}^{m}U_k(g_n)|\psi_k(0)\rangle$ (the bar above an observable quantity denotes averaging over all disorder configurations) such that the random variable $g_n$ takes the values $1$ and $0$ with probabilities (or bias) $p$ and $(1-p)$ respectively with the following effect,
\begin{equation}\label{eq:binomial}
\begin{split}
U_k(0)=e^{-iH_0(k)T},\\
U_k(1)=e^{-iH_1(k)\frac{T}{2}}e^{-iH_0(k)\frac{T}{2}}.
\end{split}
\end{equation}
In all the instances of explicit symmetry breaking (mentioned in the earlier section) realised in the Hamiltonian $H_1(k)$, the configuration averaged stroboscopic particle current and the stroboscopic change in the bulk topological invariant is observed to decay to zero for large stroboscopic intervals of observation (see Figs.~\,\ref{2b},\ref{2d}, \ref{3b} and \ref{3d} for further details).

 \section{Heat current generation through dynamical symmetry breaking through a sudden quench}

\label{sec_heat_current}

The dynamical breaking of symmetries in the SSH model is also accompanied by a non-equilibrium energy flow in the bulk which we analyse in this section. To analytically study the local energy current we resort to a local energy operator defined in the bulk\cite{zotos97}. It changes in time according to the continuity equation expressed in terms of the divergence of the heat current operator. In all the cases
discussed below, the initial Hamiltonian is the BDI SSH model in Eq.~\eqref{eq:H}. We are suddenly changing the parameters of the Hamiltonian (or including a staggered on-site potential)
to a final Hamiltonian ${\cal H}$, 
which may or may not respect the symmetries of the initial Hamiltonian.

The first situation we consider is that the {\it final post-quench} Hamiltonian ${\cal H}$  does not have a onsite potential and respects the time reversal symmetry($\mathcal{T}$) and particle hole symmetry($\mathcal{P}$)  as given in Eq.~\eqref{eq:H} can be written as,
\begin{equation}\label{hopping}
{\cal H}=\sum_{i}{E_{i,i+1}}
\end{equation}
where the term $E_{i,i+1}$ connects the $i^{th}$ site with the $(i+1)^{th}$ site and the summation extends over the chain length assuming a periodic boundary conditions. The local energy current in the Heisenberg picture can then be written in terms of a continuity equation,
\begin{equation}\label{continuity}
\frac{\partial E_{i,i+1}^{h}(t)}{\partial t}=-i[{\cal H},E_{i,i+1}^{h}(t)]=-[j^{h}_{i+1}(t)-j^{h}_{i}(t)],
\end{equation}
where the superscripts ${h}$ imply that the corresponding operators are written in the Heisenberg picture which we shall omit in the subsequent discussion. The last equality enforces the conservation of energy thereby defining the local energy current operator $j^E_{i}$, where the superscript implies the heat current. Given the form of the Hamiltonian Eq.~\eqref{hopping}, it is straightforward to show that the heat current operator assumes the form,
\begin{equation}\label{jh}
j_i^E=-i[E_{i-1,i},E_{i,i+1}].
\end{equation}
Comparing with the SSH Hamiltonian Eq.~\eqref{eq:H} and recasting it to a modified form,
\begin{multline}\label{Heo}
{\cal H}=\sum_{i}{(vc^{\dagger}_{2i-1}c_{2i}+wc^{\dagger}_{2i}c_{2i+1}+h.c.)}=\\
\sum_{i}{(E_{2i-1,2i}+E_{2i,2i+1})}
\end{multline}
where the odd sites reside on the A sublattice and the even sites reside on the B sublattice.
The total energy current operator can then be formulated by summing over all the even and odd sites of the chain and using the anti-commutation relations in Eq.~\eqref{eq:anticom},
\begin{multline}\label{j1i}
J^E=-i\sum_{i}vw\left(c^{\dagger}_{2i-2}c_{2i}+c^{\dagger}_{2i-1}c_{2i+1}\right)+\\
v^*w^*\left(c_{2i-2}c^{\dagger}_{2i}+c_{2i-1}c^{\dagger}_{2i+1}\right).
\end{multline}
Now, utilizing the translational invariance of the periodically wrapped chain, one can rewrite the energy current operator in Fourier space explicitly reintroducing the sublattice index ($A$ or $B$),
\begin{equation}\label{j1ka}
	J^E_k=\sum_{k}{\left(c^{\dagger}_{kA}~c^{\dagger}_{kB} \right)j^E_k\begin{pmatrix}
	c_{kA}\\
	c_{kB}
	\end{pmatrix}},
\end{equation} 
where $j^E_k$ is a $2\times2$ matrix,
\begin{equation}\label{j1kb}
j^E_k=-i\left(vw e^{-ik}-v^*w^*e^{ik} \right)\mathbb{I},
\end{equation}
$\mathbb{I}$ being a $2\times2$ identity matrix. The total energy current in a state $\ket{\psi_k}$ is then obtained by integrating the expectation value of $J^E_k$ over the complete Brillouin zone,
\begin{equation}\label{j1}
J^E_{\psi}=\int_{BZ}dk \bra{\psi_k}j_k^E\ket{\psi_k},
\end{equation}
which in the SSH model sums up to,
\begin{equation}\label{j1b}
J^E_{\psi}=2\int_{-\pi}^{\pi}dk~\operatorname{Im}\left( vwe^{-ik}\right).
\end{equation}
It is important to note that the energy current operator $j^E_k$ is a multiple of identity and hence, commutes with the Hamiltonian (as is clear from Eq.~\eqref{j1kb}). This causes $J^E_{\psi}$ to remain invariant in time. Utilizing the above analytical framework, we study the behaviour of the heat current in the light of different symmetries of the evolving Hamiltonian following a sudden quench. \\

\subsection{Preserving both $\mathcal{P}$ and $\mathcal{T}$}
 We consider a sudden quench of the hopping strengths from an initial value of $v$ and $w$ to $v^{\prime}$ and $w^{\prime}$ respectively such that it preserves both the symmetries ($\mathcal{P}$ and $\mathcal{T}$) in the final Hamiltonian (refer to Eq.~\eqref{eq:H}). For the total bulk energy current, one obtains the expression,
\begin{equation}\label{je1}
J^E_{\psi}=2v^{\prime}w^{\prime}\int_{-\pi}^{\pi}dk~\sin{k}=0,
\end{equation}
Also, since the energy current is a conserved quantity in this case, it remains zero throughout the time evolution. 

\subsection{Breaking both $\mathcal{P}$ and $\mathcal{T}$}
When the inter-sublattice hopping parameter i.e. $w$  is suddenly changed to  a complex value in the final evolving Hamiltonian 
 both $\mathcal{P}$ and $\mathcal{T}$ are broken.  From the expression for the energy current in Eq.~\eqref{j1b}, it is straightforward to show that the current operator still remains a multiple of $\mathbb{I}$
 resulting in 
\begin{equation}
J^E_{\psi}=2v^{\prime}w^{\prime}\int_{-\pi}^{\pi}dk~\cos{k}=0
\end{equation} 
Hence in this case as well, the bulk heat current vanishes for all times. This is to be contrasted to the case of a surviving time dependent particle current as has been illustrated in Fig.~\ref{3a}.\\

\subsection{Breaking $\mathcal{P}$ while preserving $\mathcal{T}$}
\label{sec_heat_breaking_P}

In this protocol,
 a staggered on-site potential is suddenly introduced in the free SSH chain; this breaks $\mathcal{P}$ while preserving $\mathcal{T}$ as has been shown in Eq.~\eqref{eq: B2}. Due to the  breaking of the particle hole symmetry,  the final Hamiltonian assumes the following form,
\begin{equation}\label{Hm}
{\cal H}_{\mathcal{T}}=H_{SSH}+\sum_{i}{Mc^{\dagger}_{2i}c_{2i}-Mc^{\dagger}_{2i-1}c_{2i-1}}
\end{equation}
where $H_{SSH}$ is the bare and symmetric Hamiltonian of the periodically wrapped SSH chain \eqref{Heo}. In the presence of the  onsite potential however the expression of the local energy current must be re-written to incorporate the additional diagonal terms in the Hamiltonian. Proceeding in similar lines as to the derivation of the Eq.~\eqref{jh}, one obtains,
\begin{equation}
j^E_i=-i\left(\left[E_{i-1,i},E_{i,i+1}\right]+\left[E_{i-1,i},E_{i,i}\right]\right)
\end{equation}
where $E_{i,i}$ are the symmetry breaking diagonal terms of the Hamiltonian $H_{\mathcal{T}}$. 
Simplifying the above expression using the post quench Hamiltonian $H_{\mathcal{T}}$, the fermion anti-commutations relations and the translational invariance of the chain, the local current operator is expressed in the momentum space as,
\begin{equation}\label{Jmk}
J^E_k=\vec{j}^E_i(k).\vec{\sigma}
\end{equation}
where,
\begin{eqnarray}\label{jmk}
\nonumber j_0^E(k)=2\operatorname{Im}\left(vwe^{-ik}\right)\\
\nonumber j^E_x(k)=Mw\sin{k}\\
\nonumber j^E_y(k)=M(v+w\sin{k})\\
 j_z^E(k)=0.
\end{eqnarray}
Interestengly, it is observed that apart from the contribution proportional to the identity matrix (as in Eq.~\eqref{j1kb}), nontrivial non-diagonal terms have appeared in the local heat current operator in the presence of the stagerred on-site potential.

Using the above components to evaluate the total heat current according to Eq.~\eqref{j1}, one obtains the following analytic expression for the heat current,
\begin{multline}
J^E(t)=M\int_{-\pi}^{\pi}dk~\frac{\sin{(2m_ft)}}{m_f}[j^E_y(k)\cos{\phi}-j_x^E(k)\sin{\phi}]
\end{multline}
where $H_{\mathcal{T}(k)}=\vec{m}^f(k).\vec{\sigma}$, $m_f=|\vec{m}_f(k)|$ $H_{SSH}(k)=\vec{m}^i(k).\vec{\sigma}$ and $\phi=\tan^{-1}{\left[\frac{m_y^i(k)}{m_x^i(k)}\right]}$. Thus, the SSH model now shows non-zero flow of heat (see Fig.~\ref{4a}) in the bulk as a consequence of the dynamical breaking of $\mathcal{P}$ symmetry while the $\mathcal{T}$ symmetry remains intact.\\

\subsection{Breaking $\mathcal{T}$ while preserving $\mathcal{P}$}
\label{sec_heat_breaking_T}

 It is also possible to break the $\mathcal{T}$ symmetry of the SSH model while keeping the $\mathcal{P}$ symmetry intact by suddenly switching on a complex staggered next nearest neighbour hopping term which renders the single particle final Hamiltonian $H_{\mathcal{P}}(k)$ to be of the form, $H_{\mathcal{P}}(k)=\vec{m}^f(k).\vec{\sigma}$ such that $(m_x^f,m_y^f,m_z^f)$ are (even, odd, odd) functions of $k$ respectively. Now,
\begin{equation}\label{hP}
{\cal H}_{\mathcal{P}}=H_{SSH}+\sum_{i}{\left( fc^{\dagger}_{2i}c_{2i+2}+fc^{\dagger}_{2i-1}c_{2i+1}+h.c.\right)}.
\end{equation}
Setting the next nearest hopping strength to be complex, $f=\lambda i\text{~where~} \lambda\in\mathbb{R}$ yields the final Hamiltonian,
\begin{equation}\label{hkP}
{\cal H}_{\mathcal{P}}(k)=(v+w\cos{k})\sigma_x+w\sin{k}\sigma_y+\lambda\sin{k}\sigma_z
\end{equation}

 This clearly shows that the $\mathcal{T}$ symmetry has been broken in the system while keeping $\mathcal{P}$ preserved throughout. As a result, the energy current however will now have two contributions, one from the nearest neighbour hopping and the other from the next nearest neighbour hopping originating from the same site, i.e.
 \begin{multline}\label{jep1}
 j^E(k)=\sum j^{(1)}(k)_{A}+j^{(1)}(k)_{B}\\
 +\sum j^{(2)}(k)_{A}+j^{(2)}(k)_{B}
 \end{multline}
 where the summation extends over all the lattice sites.  $j^{(1)}(k)$ are the nearest neighbour current and $j^{(2)}(k)$ are the next nearest neighbour current. Expressing the total heat current operator in the $2\times 2$ sublattice basis as Eq.~\eqref{Jmk}, one obtains,
\begin{eqnarray}\label{jfk}
\nonumber j_0^E(k)=2\operatorname{Im}\left(vwe^{-ik}-\lambda^2 e^{-2ik}\right)\\
\nonumber j^E_x(k)=-\lambda\left(w-v\cos{k}\right)\\
\nonumber j^E_y(k)=-\lambda\left(v\sin{k}\right)\\
j_z^E(k)=0,
\end{eqnarray}
which is non-zero when integrated over the complete Brillouin zone (Fig.~\ref{4b}). Thus, we see that the heat current in such a situation is non-zero although the particle current vanishes when $\mathcal{P}$ is preserved by the final Hamiltonian.

\section{Conclusions}
In this work, we  have studied the particle and heat transport properties of a time-dependent 1D topological quantum system. The goal is to test the robustness of the 1D topological phase against the inclusion of dynamical perturbations  and the possible change in the associated  winding number. Therefore, we resort to the simple 1D SSH model, to investigate the effect on the  transport, namely the polarisation current
and the heat current, in such systems when the symmetries of the underlying system may be broken by the time-dependent perturbations.
 We focus on the issue whether the winding number can be changed through out of equilibrium drives and whether such a change can be captured in the transport properties of the system.  In our work, the time evolution of the initial state of the 1D system is introduced through  time-periodic drives, quantum quenches and noisy perturbations that break the perfect time periodicity. We see that through the dynamical breaking of certain discrete (non-crystal) symmetries namely the particle-hole ($\mathcal{P}$), time-reversal ($\mathcal{T}$) and the chiral symmetry ($\mathcal{S}$), there can be a generation of either particle or heat current in the bulk of the 1D chain accompanied by a change in the winding invariant with time.\\
 
  Specifically in the perfectly periodic situation, we observe the following behaviour: (1)  when only the particle-hole symmetry is broken in the instantaneous Hamiltonian  or within the period of a drive, particle current is generated. However, the breaking of this  $\mathcal{P}$ in the instantaneous Hamiltonian does not guarantee that the bulk Floquet Hamiltonian, which governs the dynamics of the system at stroboscopic intervals, will also have a broken $\mathcal{P}$ symmetry. Nonetheless, if the $\mathcal{P}$ symmetry still  remains   preserved in the Floquet Hamiltonian, the winding number will still be conserved when observed stroboscopically. (2) When the Floquet Hamiltonian also breaks the $\mathcal{P}$ symmetry, we see the stroboscopic generation of a particle current in the system even when the initial state carried zero current. (3) The generated current in case (2) following some initial transients eventually settles down to a steady non-zero value asymptotically in time. \\
 We next subject the system to biased random noisy perturbations that break the perfect periodicity of the drive. Interestingly,  when the perfect time periodicity within a period is broken due to the presence of such perturbations, the particle current although shows a significant pre-thermal value, eventually decays to zero asymptotically with time reflecting the fact that the system reaches an infinite temperature ensemble. We note that this happens even when both the drive and the noisy perturbations break the $\mathcal{P}$ symmetry explicitly. \\
 
 Finally, we also probe the out of equilibrium behavior of energy transport in the bulk of the system due to time-dependent driving in the form of sudden quenches. We observe that
   even when there is no heat current flowing in the system initially, dynamical breaking of either $\mathcal{P}$ or $\mathcal{T}$, but not both,  results in the generation of a heat current in the bulk of the system. This is notably different in comparison to the dynamical conditions that result in the flow of a particle current in the system.\\

%
   
  The  polarisation current as well as the energy current flowing through the bulk of the chain being observables   can be experimentally measured in a transport set up and hence the predictions
  made in this work can be verified. We recall that  in the process of an adiabatic quantum pump charactarized by the topology of the pumping cycle, the dynamical state of the system follows the adiabatically evolving Hamiltonian. Therefore,  the conclusions  reached
   through our work regarding the symmetries of the Floquet Hamiltonian, would naturally manifest in the topological transport of charge across a SSH chain under an adiabatic periodic perturbation.

\begin{acknowledgments}
We acknowledge  for Sourav Bhattacharjee, Sudarshana Laha and Somnath Maity for  discussions. SB acknowledges CSIR, India for financial support.
\end{acknowledgments}

\end{document}